\documentclass{PoS}

\def\a{{\alpha}}
\def\b{{\beta}}
\def\d{{\delta}}
\def\D{{\Delta}}
\def\e{{\epsilon}}
\def\g{{\gamma}}

\def\k{{\kappa}}

\def\L{{\Lambda}}

\def\t{{\tau}}

\def\mc#1{{\mathcal #1}}

\def\eqref#1{{(\ref{#1})}}

\title{On the Cottingham formula and the electromagnetic contribution to the proton-neutron mass splitting}

\ShortTitle{Cottingham and $M_p-M_n$}

\author{\speaker{Andr\'{e} Walker-Loud}\thanks{I am indebted to my colleagues, Carl Carlson and Jerry Miller with whom this work was possible.
}\\
        Nuclear Science Division, Lawrence Berkeley National Laboratory, Berkeley, CA 94720, USA\\
        E-mail: \email{walkloud@lbl.gov}}

\abstract{
The excess mass of the neutron over the proton arises from two sources within the Standard Model, electromagnetism and the splitting of the down and up quark masses.  The Cottingham Formula provides a means of determining the QED corrections from the forward Compton Amplitude, but this is challenged by the need for a subtraction function and the mixing of the QED and QCD (electro-weak) effects.  
I review the present understanding of the Cottingham Formula, including a discussion on the development of the formula, its renormalization which induces the mixing of QED and QCD effects, and the necessary modeling of the subtraction function that must be done to arrive a numerical prediction.
I summarize the Regge Model originally proposed by Gasser and Leutwyler
and
I also review the proposed model by Walker-Loud, Carlson and Miller, which is an interpolation function between the low and high $Q^2$ regimes, both of which are anchored by rigorous theoretical underpinnings, for which I argue a more reliable theoretical uncertainty estimate can be obtained.
}

\FullConference{The 9th International workshop on Chiral Dynamics\\
		17-21 September 2018\\
		Durham, NC, USA}

\begin{document}

\section{The Cottingham Formula}
The Cottingham Formula~\cite{Cottingham:1963zz} provides the following, model independent, leading order QED mass shift for the isovector nucleon mass, under one assumption:%
\footnote{As we will explain in some detail, the one requisite assumption is that, in the fixed $Q^2$ dispersive representation of the scalar functions arising in the Compton Amplitude, the $T_2(\nu,Q^2)$ function does not require a subtracted dispersion integral while the $T_1(\nu,Q^2)$ function only requires a single subtraction.}
\begin{eqnarray}
\label{eq:Cottingham}
\delta M^\gamma &=& \frac{\a_{fs}}{2\pi M}\int_0^{\L^2} \hspace{-4pt} dQ^2 \Bigg\{
	\frac{1}{2(1+\t_{el})} \left[
	(G_E^2 -2\t_{el}G_M^2)\left(
		\frac{1}{\sqrt{\t_{el}}}(1+\t_{el})^{3/2}-\t_{el}
	\right)
	+3\t_{el}G_M^2
	\right]
\nonumber\\&&\qquad
	+\int_{\nu_{th}}^\infty \hspace{-8pt} d\nu \left[
	F_1 \frac{3 Q}{\nu^2} \left( 
		\t^{3/2} -\t\sqrt{1+\t} +\frac{\sqrt{\t}}{2}
	\right)
	+F_2\frac{M}{Q\nu} \left(
		(1+\t)^{3/2} -\t^{3/2} -\frac{3}{2}\sqrt{\t}
	\right)
	\right]
\nonumber\\&&\qquad
	-\frac{3}{8} T_1^{inel}(0,Q^2)
	\Bigg\}
\nonumber\\&&
	\pm\frac{3\a_{fs}}{8\pi M}\ln \left(\frac{\L_{\rm UV}^2}{\L^2}\right)
		\frac{e_u^2 m_u - e_d^2 m_d}{\d}
		\langle p| \d(\bar{u}u -\bar{d}d)|p\rangle\, .
\end{eqnarray}
In this expression, 
$\a_{fs} = \frac{e^2}{4\pi}$,
$M$ is the nucleon mass,
$\t_{el} = \frac{Q^2}{4M^2}$, $\t = \frac{\nu^2}{Q^2}$,
$\nu_{th} = m_\pi + \frac{m_\pi^2 + Q^2}{2M}$ is the inelastic threshold where a real pion can be produced,
$G_{E,M}(Q^2)$ are the elastic electric and magnetic form factors,
$F_{1,2}(\nu,Q^2)$ are the inelastic structure functions and
$T_1^{inel}(0,Q^2)$ is a subtraction function arising from the dispersive representation of the inelastic $T_1$ structure function.
The final correction provided as $\pm$ is a theoretical uncertainty arising from estimating finite corrections arising from the renormalization procedure.
The counterterm needed for the renormalization is proportional to the isovector  quark mass operator, $\langle p| \d(\bar{u}u -\bar{d}d)|p\rangle$, with $2\d\equiv(m_d-m_u)$ where $m_{u,d}$ are the $up$ and $down$ quark masses and $e_{u,d}$ are the electromagnetic charges of the quarks.
As discussed by Collins~\cite{Collins:1978hi}, this expression has been renormalized such that the UV divergence, when $\L_{\rm UV}\rightarrow\infty$, has been exactly cancelled by the counterterm.
Given the non-perturbative nature of QCD, the only contribution from the counterterm we can determine exactly, without resorting to non-perturbative methods such as lattice QCD (LQCD), is the exact cancellation of the UV divergence appearing from the integral and the isovector quark mass operator.
We are left with unknown finite contributions arising from the non-asymptotic scaling regime.
However, these corrections can be estimated following arguments very similar to naive dimensional analysis~\cite{Manohar:1983md} by varying the asymptotic QCD scale, $\L_{\rm UV}$ and the scale at which the renormalization was applied, $\L$, which must still be in the perturbative region to ensure this finite ambiguity remains small.%
\footnote{In the full electroweak standard model, the quarks are massless, so there are no operators to serve as counterterms and thus there can not be any UV divergence.  The divergence is in fact cancelled by an opposite contribution arising from a virtual Z-boson loop~\cite{Weinberg:1972ws}, which mimics a Pauli-Villars regulator of the photon propagator.  This does not remove the ambiguity of small finite corrections which arise from the cancellation of the photon and Z-boson corrections.} 

The derivation of this formula is straightforward.
One begins by convolving a photon propagator with the forward Compton Amplitude (and regulating the integral and adding a counterterm)
\begin{equation}
\label{eq:dM_qed}
	\delta M^\gamma = \frac{i}{2M} \frac{\a_{fs}}{(2\pi)^3}
\int_R d^4 q
\frac{T_{\mu\nu}(p\cdot q,q^2)g^{\mu\nu}}{q^2 +i\epsilon} + c.t.(R)\, ,
\end{equation}
where we have left the regulator/renormalization-scheme (R) implicit.
The Compton Amplitude
\begin{equation}
\label{eq:compton}
T_{\mu\nu} = \frac{i}{2} \sum_{s} \int d^4 x e^{i q\cdot x}
	\langle p, s| \mathrm{T} \{ J_\mu(x) J_\nu(0) \} | p, s\rangle\, ,
\end{equation}
is given by the time-ordered product of two electromagnetic currents between on-shell proton states and summed over the spin orientations.
The Cottingham Formula, Eq.~\eqref{eq:Cottingham} is obtained from Eq.~\eqref{eq:dM_qed} after straightforward manipulations of the integrand followed by a dispersive parameterization of the structure functions arising in the Compton Amplitude, Eq.~\eqref{eq:compton},
\begin{eqnarray}
\d M^\g = \frac{\a_{fs}}{8\pi^2} \int_0^{\L^2} \hspace{-8pt} dQ^2
	\int_{-Q}^{Q} \hspace{-8pt} d\nu \frac{\sqrt{Q^2 - \nu^2}}{MQ^2}
	\left[
	-3T_1(i\nu,Q^2) + \left(1 - \frac{\nu^2}{Q^2} \right) T_2(i\nu,Q^2)
	\right]
	+c.t.(\L)\, .
\end{eqnarray}

The derivation of this formula arose from a community desire to understand the mass splitting between the neutron and proton.
The naive expectation was that the electromagnetic self-energy of the proton would be larger than that of the neutron given their electric charges.
However, it was understood that there was a near perfect isospin symmetry in the strong interactions~\cite{Heisenberg1932} and that the nucleons were not fundamental, but of a composite nature~\cite{Chambers:1956zz,Yearian:1958shh}.
This led to the serious speculation that the electromagnetic self-energy of the neutron may be greater than the proton due to contributions from high energies in the integral, or deep inside the nucleons~\cite{Feynman:1954xxl}.

This idea was formalised with dispersion theory~\cite{Cini:1959szx} where it was noted that, if, instead of cutting off the integral, one used the measured elastic form-factors of the nucleon, it was impossible this relation would yield even the correct sign for the nucleon mass splitting.  
This idea was followed up~\cite{Pande1962} with a more recent measurements and parameterization of the form factors~\cite{PhysRevLett.8.381}, using the \textit{Clementel-Villi} form~\cite{Hofstadter:1961zz,Bergia:1961zz}.  These elastic form factors included constant terms which lead to divergences when integrated over $Q^2$.  With suitable choices of regulator scale~\cite{Pande1962}, similar to those chosen in Ref.~\cite{Feynman:1954xxl}, again, it seemed the QED corrections could give rise to a mass excess of the neutron over the proton without any other sources of isospin violation.

These works had considered only the elastic nucleon structure.  Soon after this, Cottingham provided a complete expression (except for the renormalization), including contributions from the inelastic structure of the nucleons, noting that it was possible these contributions could be significant, but that experimental knowledge at the time was prohibitive from determining their contribution~\cite{Cottingham:1963zz}.
Cottingham also noted that one of the two structure functions may require a subtraction in the dispersive representation, which would invalidate estimates to date.
Shortly after this work, Harari used ``the most successful and least controversial prediction of Regge pole theory'' to conclusively demonstrate that a subtracted dispersion integral was necessary as even in the isovector combination, one of the stucture functions scales as $|t_1^{\D I=1}(\nu,Q^2)|_{\nu\rightarrow \infty} \propto \nu^{0.4}$ at fixed $Q^2$~\cite{Harari:1966mu}.  Therefore, in Cauchy's contour integral formula, the contribution from the infinite arc is not small enough to vanish as $\nu\rightarrow\infty$.
Harari noted that the subtraction term could be related to the ``tadpole'' operators introduced by Coleman and Glashow~\cite{Coleman:1963pj} to explain the observed mass splittings in the baryon octet.
In Ref.~\cite{Gasser:1974wd}, Gasser and Leutwyler proposed a model of the subtraction function based on Regge Theory which they used to estimate the isovector mass correction~\cite{Gasser:1982ap}.

There has been a recent renewed interest in understanding the QED correction to $M_p-M_n$~\cite{WalkerLoud:2012bg,WalkerLoud:2012en,Walker-Loud:2013yua,Walker-Loud:2014iea,Thomas:2014dxa,Erben:2014hza,Gasser:2015dwa,Leutwyler:2015jga,Tomalak:2018dho}, originating from a proposal to use the nucleon mass splitting as an alternative means to determine $m_d - m_u$ with LQCD~\cite{WalkerLoud:2010qq}.
However, the present theoretical uncertainty arising from the Cottingham Formula is not competitive with that from LQCD calculations which either compute only the strong $m_d-m_u$ correction, and then determine the QED correction by subtracting this from the experimental value~\cite{Beane:2006fk,WalkerLoud:2009nf,deDivitiis:2011eh,Horsley:2012fw,deDivitiis:2013xla,Brantley:2016our,Heffernan:2017hwa}, or from calculations which incorporate QED in the LQCD calculations~\cite{Blum:2010ym,Borsanyi:2013lga,Borsanyi:2014jba,Horsley:2015eaa}.

\section{Dispersive Representation of the Compton Amplitude \label{sec:subtraction}}
There are two common parameterizations of the Compton Amplitude of the nucleon
\begin{eqnarray}
\label{eq:little_t_big_T}
T_{\mu\nu} = 
	d_{\mu\nu}^{(1)}\ q^2\ t_1(\nu,q^2)
	-d_{\mu\nu}^{(2)}\ q^2\ t_2(\nu,q^2)
	= -D_{\mu\nu}^{(1)}\ T_1(\nu,q^2)
	+D_{\mu\nu}^{(2)}\ T_2(\nu,q^2)\, ,
\end{eqnarray}
where
\begin{eqnarray}
d_{\mu\nu}^{(1)} = D_{\mu\nu}^{(1)} &=& g_{\mu\nu} - \frac{q_\mu q_\nu}{q^2}\, ,
\nonumber\\
d_{\mu\nu}^{(2)} &=& \frac{1}{M^2}\left(
	p_\mu p_\nu 
	- \frac{p\cdot q}{q^2} (p_\mu q_\nu + p_\nu q_\mu)
	+ \frac{(p\cdot q)^2}{q^2} g_{\mu\nu}
	\right)\, ,
\nonumber\\
D_{\mu\nu}^{(2)} &=& \frac{1}{M^2}
	\left( p_\mu - \frac{p\cdot q}{q^2} q_\mu \right)
	\left( p_\nu - \frac{p\cdot q}{q^2} q_\nu \right)\, ,
\end{eqnarray}
leading to the relation between the scalar functions
\begin{equation}
T_1(\nu,q^2) = -q^2 t_1(\nu,q^2) + \nu^2 t_2(\nu,q^2)\, ,
\qquad
T_2(\nu,q^2) = -q^2 t_2(\nu,q^2)\, .
\end{equation}
There is an advantage to choosing the ``little-$t$'' parameterization (as we refer to it) versus the ``big-$T$'' parameterization.
In the deep inelastic scaling (DIS) regime, the imaginary part of $t_1$ is the Callan-Gross function~\cite{Callan:1969uq} which vanishes for $\lim {Q^2\rightarrow\infty}$ and fixed Bjorken-$x = \frac{Q^2}{2M\nu}$,
\begin{equation}
2{\rm Im}\, t_1(\nu,Q^2) = 
	\frac{2\pi M \nu}{Q^4} \left[ 2x F_1(x,Q^2) - F_2(x,Q^2) \right]\, .
\end{equation}
For a point particle, therefore, $t_1(\nu,Q^2) = 0$ exactly at leading order in QED.

Regardless of the parameterization, it is unambiguously accepted that the $t_1$ ($T_1$) function requires a subtraction in the dispersive representation while the $t_2$ ($T_2$) function does not.  The $t_i$ functions are crossing-symmetric, $t_i(-\nu,Q^2) = t_i(\nu,Q^2)$ and so the fixed-$Q^2$ dispersive representations are given by
\begin{eqnarray}
T_1(\nu,Q^2) &=& T_1(0,Q^2) 
	+ \frac{\nu^2}{2\pi} \int_{\nu_{th}}^\infty \hspace{-8pt} d\mu
	\frac{2\mu}{\mu^2 (\mu^2 - \nu^2)} 2{\rm Im}\, T_1(\mu+i\e,Q^2)\, ,
\nonumber\\
T_2(\nu,Q^2) &=& \frac{1}{2\pi} \int_{\nu_{th}}^\infty \hspace{-8pt} d\mu
	\frac{2\mu}{\mu^2 - \nu^2} 2{\rm Im}\, T_2(\mu+i\e,Q^2)\, .
\end{eqnarray}
If the subtraction is made at $\nu=0$, then the subtraction function arising from the little- and big-T representations, Eq.~\eqref{eq:little_t_big_T} are identical and both proportional to $T_1(0,Q^2)$.
The major challenge in making a precise determination of the isovector QED mass correction is that while the imaginary parts of the structure functions can be related to experimentally measured cross sections (with the normalization in Eq.~\eqref{eq:compton})
\begin{equation}
2{\rm Im}\, T_1(\nu+i\e,Q^2) = 4\pi F_1(\nu,Q^2)\, ,
\qquad
2{\rm Im}\, T_2(\nu+i\e,Q^2) = 4\pi \frac{M}{\nu}F_2(\nu,Q^2)\, ,
\end{equation}
the subtraction function, $T_1(0,Q^2)$, can not be directly related to measured quantities.
We next discuss the renormalization followed by strategies  to deal with the subtraction function.

\subsection{Subtraction function and renormalization \label{sec:renormalization}} 
Collins first worked out the renormalization of the Cottingham formula~\cite{Collins:1978hi}, which we summarize here.
It is instructive to recall that the divergence that appears in the Cottingham Formula and the renormalization must behave just like the QED self-energy correction to the the electron mass, as ultimately, at asymptotically high $Q^2$, the divergence will arise from the QED self-energy shift of the individual quarks.
If the regularization scheme used respects chiral symmetry, then the counterterm must be proportional to a quark mass operator.%
\footnote{If the self-energy of an individual nucleon is considered, there are additional counterterms that can arise, but in the isovector case, the only operator with the required flavor structure is the isovector quark mass operator.} 
If we focus on the DIS contribution to the Cottingham Formula, we have
\begin{equation}
\label{eq:dm_dis}
\d M_{DIS} \simeq \frac{3\a_{fs}}{16\pi M} \int_{\L_{DIS}}^{\L_{UV}}
	\frac{dQ^2}{Q^2} \left\{
	-Q^2 T_1(0,Q^2) 
	+ M^2 \int_0^1 \hspace{-6pt}dx \left[2x F_1(x) + F_2(x)\right]
	\right\}
	+c.t.
\end{equation}
From the Operator Product Expansion, the asymptotic behavior of these terms is
\begin{eqnarray}
-Q^2 T_1(0,Q^2) &\sim& 
	-Q^2 \sum_i \left[ 
		\frac{1}{4}M^2 (Q^2 C_{1,i}^2 + C_{2,i}^2) \langle \mc{O}^{i,2}\rangle			+C_{1,i}^0 \langle \mc{O}^{i,0}\rangle
	\right]\, ,
\nonumber\\
M^2 \int_0^1 \hspace{-6pt}dx \left[ 2x F_1(x) +F_2(x)\right] &\sim&
	\phantom{-}Q^2\sum_i \phantom{\bigg[}
	\frac{1}{4}M^2  (Q^2 C_{1,i}^2 + C_{2,i}^2) \langle \mc{O}^{i,2}\rangle\, .
\end{eqnarray}
One observes that the divergences arising from the structure functions appearing in the dispersive integrals are exactly cancelled by terms which arise in the subtraction function.
Therefore, the equal and opposite UV divergences arising from the counterterm are intimately connected to the subtraction function:
\begin{equation}
\d M_{DIS} \simeq \frac{3\a_{fs}}{16\pi M} \int_{\L_{DIS}}^{\L_{UV}}
	dQ^2 \sum_i (-C_{1,i}^0)
	+c.t.\, ,
\qquad
C_{1,i}^0 \propto \frac{1}{Q^2}\, .
\end{equation}
As the short distance operator scales as $C_{1,i}^0 \propto \frac{1}{Q^2}$, we see there is a $\ln(Q^2)$ divergence, as we expect, which is exactly cancelled by the counterterm (the divergence associated with $\L_{UV}\rightarrow\infty$ is exactly cancelled).
For large values of $\L_{DIS} < \L_{UV}$ the cancellation is not exact, but finite corrections are small, controlled by perturbative QCD.
Using arguments similar to naive dimensional analysis~\cite{Manohar:1983md}, we can estimate the size of these finite corrections by varying the scale at which the renormalization is performed, which in the above expression, is given by $\L_{DIS}$.
In the present case, we even know the value of the short distance operator as it can be determined by computing the Wilson coefficients~\cite{Collins:1978hi}.%
\footnote{The error in Ref.~\cite{Collins:1978hi} reported by Hill and Paz~\cite{Hill:2016bjv} does not impact the isovector nucleon mass renormalization.} 
For $M_p-M_n$, we can estimate the finite, residual corrections from the inexact cancellation of the counterterm and the structure functions
\begin{equation}
\label{eq:dM_residual}
\d M_{residual} = \pm\frac{3\a_{fs}}{8\pi M} \int_{\L^2}^{\L_{\rm UV}^2}
	\frac{dQ^2}{Q^2}\frac{e_u^2 m_u - e_d^2 m_d}{\d} 
	\langle p|\d (\bar{u}u - \bar{d}d)|p\rangle\, .
\end{equation}
To arrive at this expression, we used the approximate isospin symmetry of QCD to relate matrix elements such as $\langle n| \bar{u}u|n\rangle \simeq \langle p|\bar{d} d|p\rangle$, see for example~\cite{Miller:1997ya,Wagman:2014nfa}

It is interesting to note, the renormalization of the Cottingham Formula causes a mixing of QED and QCD ($m_d-m_u$) effects, such that a rigorous separation of the contribution to $M_p-M_n$ from QED and QCD is not possible in a scheme independent  manner.
However, as Colins noted, the size of the isovector quark mass splitting is comparable to the quark masses themselves, at least when compared to the typical QCD scale, and so, practically speaking, this mixing of QED and QCD is numerically in size the same as second order isospin breaking corrections.  Therefore, if we are content with understanding such isospin breaking effects to leading order, we can meaningfully speak of separating the QCD and QED corrections.
A recent proposal for incorporating QED with LQCD calculations with a hadronic scheme is constructed to respect this LO separability~\cite{Bussone:2018ybs}.

\subsection{Parameterizing the Subtraction Function} 
Given this renormalization, we now have a complete expression for the leading QED corrections to $M_p-M_n$, Eq.~\eqref{eq:Cottingham}.
In order to make a prediction for this value, we must figure out how to handle the unknown subtraction function contribution, the integral over $T_1(0,Q^2)$.
As this function can not be directly related to measured cross sections, it is not possible determine the mass correction without introducing a model for $T_1(0,Q^2)$.
In the literature, there are essentially two models proposed, the original model of Gasser and Leutwyler~\cite{Gasser:1974wd} and the more recent one proposed by us~\cite{WalkerLoud:2012bg}.  We comment on these approaches in the following sections.

\subsubsection{Regge Model}
Gasser and Leutwyler (GL)~\cite{Gasser:1974wd} and Gasser, Hoferichter, Leutwyler and Rusetsky (GHLR)~\cite{Gasser:2015dwa} make the ansatz that the $t_1(\nu,Q^2)$ function can be separated as
\begin{equation}
t_1(\nu,q^2) = \bar{t}_1(\nu,q^2) + t_1^R(\nu,q^2)\, ,
\end{equation}
where $\bar{t}_1(\nu,q^2)$ satisfies an unsubtracted fixed-$q^2$ dispersion relation while the Regge contribution requires a subtraction.  They propose a  model of the Regge contribution
\begin{equation}
t_1^R(\nu,q^2) = -\sum_{\a>0} \frac{\pi \b_\a(Q^2)}{\sin \pi \a}\left[
	(s_0 -s_+ -i\e)^\a + (s_0 -s_--i\e)^\a
	\right]\, ,
\end{equation}
where $s_0\geq M^2$ is a constant and $s_\pm = M^2 \pm 2M\nu-Q^2$.

The main shortcoming of this model is that the Regge region is precisely where we do not have any control over the theoretical uncertainties arising from QCD as it is neither a region where perturbative QCD applies nor is there an effective field theory description.  This makes it impossible to provide a rigorous estimate of the theoretical uncertainty associated with this ansatz.

A further shortcoming is that, for reasons not entirely clear to us, despite our significantly improved understanding of nucleon structure, GHLR did not provide a prediction for the QED correction to $M_p-M_n$, but simply referred to the old GL result of $0.76(30)$~MeV from 1975.

\subsubsection{Interpolating function}

In Ref.~\cite{WalkerLoud:2012bg}, we proposed to rely on our rigorous theoretical understanding in the low and high $Q^2$ regimes and only model the interpolation between them.
While it remains impossible to provide a completely systematically improvable theoretical uncertainty for $T_1(0,Q^2)$ over the entire region of $Q^2$ necessary, under the assumption that $T_1(0,Q^2)$ is a smooth function between the low and high $Q^2$ regions, we can make a reasonable estimate for the upper bound on the size of the error we are making with this model.
In the high $Q^2$ regime, perturbative QCD requires~\cite{Collins:1978hi}
\begin{equation}
	T_1(0,Q^2) \propto Q^{-2}\, ,\qquad \textrm{for $Q^2 \gg \L_{QCD}^2$.}
\end{equation}
In the low $Q^2$ regime, effective field theory constrains the behavior~\cite{Pachucki:1996zza,Pineda:2002as,Pineda:2004mx,Carlson:2011zd,Hill:2011wy,Griesshammer:2012we,Birse:2012eb}
\begin{eqnarray}
\label{eq:T1_low_energy}
T_1(0,Q^2) &=& 2\k(2+\k) 
	- Q^2\left\{
		\frac{2}{3}\left[(1+\k)^2 r_M^2 - r_E^2\right] 
		+\frac{\k}{M^2}
		-2M \frac{\b_M}{\a_{fs}}
	\right\}
\nonumber\\&&
	+Q^4 \frac{2}{3} \frac{g_A^2 \k_s}{(4\pi F_\pi m_\pi)^2}
	+\mc{O}(Q^6)\, .
\end{eqnarray}
$\k$, $r_{E,M}$ and $\b_M$ are the anomalous magnetic moment, electric and magnetic charge radii and the magnetic polarizability respectively.
The $\mc{O}(Q^4)$, proportional to the isoscalar $\k_s$, is specific to the isovector case with the full expression in Ref.~\cite{Birse:2012eb}.
Eq.~\eqref{eq:Cottingham} depends upon $T_1^{inel.}(0,Q^2)$, which is given Eq.~\eqref{eq:T1_low_energy} without any of the terms which would arise from expanding elastic form factors:
\begin{equation}
\label{eq:T1_low_inel}
T_1^{inel.}(0,Q^2) = Q^2 2M\frac{\b_M}{\a_{fs}}
	+Q^4 \frac{2}{3} \frac{g_A^2 \k_s}{(4\pi F_\pi m_\pi)^2}
	+\cdots\, .
\end{equation}
Using this equation directly in Eq.~\eqref{eq:Cottingham} leads to power divergent integral.  Since we know the asymptotic behavior must scale like $Q^{-2}$, it is natural to re-sum these corrections into a form factor, which is where the dominant model uncertainty arises.
The most significant challenge in applying this interpolating model to make a prediction is that it is extremely challenging to isolate the isovector magnetic polarizability~\cite{Griesshammer:2012we}.
Until this quantity can be determined with a definite sign, it is not worth improving the interpolating model.  We felt using the simplest form factor and leaving a very generous uncertainty was the most conservative, reasonable thing to do.  In Ref.~\cite{WalkerLoud:2012bg}, only the $Q^2$ terms were known, which led to the simplistic model
\begin{equation}
T_1^{inel}(0,Q^2) = 2M \frac{\b_M}{\a_{fs}} Q^2
	\left( \frac{m_0^2}{m_0^2 + Q^2} \right)^2\, ,
\end{equation}
with a typical dipole scale of $m_0^2\simeq0.71 \textrm{GeV}^2$.

GHLR~\cite{Gasser:2015dwa} comment that this subtraction function ``is not proportional to the masses of the two lightest quarks and can thus not be absorbed in their renormalization: the particular extrapolation proposed in [31] (here \cite{WalkerLoud:2012bg}) is not consistent with the short-distance properties of QCD.''
The first part of this sentence is of course true, but misses the point.
Following the renormalization first derived by Collins, and discussed above in Sec.~\ref{sec:renormalization}, the asymptotically short distance contribution has been exactly cancelled by the counterterm.  Therefore, the interpolating subtraction function need not match this contribution, but rather it is matching onto the difference between the counterterm contribution and the contributions from the structure functions.  The implicit idea in Ref.~\cite{WalkerLoud:2012bg} is that the integral should be cut off well before the $\ln(Q^2)$ scaling would set in, otherwise, one would be integrating into a region of $Q^2$ already accounted for in the residual contribution, Eq.~\eqref{eq:dM_residual}.
Erben, Shanahan, Thomas and Young~\cite{Erben:2014hza} appreciated this subtlety but noted that our interpolation model was $\sim400$ time larger than one would predict by extending the perturbative estimates into the hadronic regime.  They therefore suggested an improved interpolation model that more smoothly connects the hadronic and large $Q^2$ regimes, thus improving the determination:
\begin{equation}
\d M^\g_{p-n}\textrm{\cite{WalkerLoud:2012bg}} = 1.30(03)(47) \textrm{ MeV}\, ,
\qquad
\d M^\g_{p-n}\textrm{\cite{Erben:2014hza}} = 1.04(35) \textrm{ MeV}\, .
\end{equation}
The uncertainty is still dominated by lack of precise knowledge of the isovector magnetic polarizability, $\b_M^{p-n}$.

\section{Conclusions}
We have reviewed the development of the Cottingham Formula~\cite{Cottingham:1963zz}, discussed its renormalization~\cite{Collins:1978hi} and the need for introducing a model of the subtraction function so that a numerical prediction can be made.
We have commented on the Regge Model originally proposed by Gasser and Leutwyler~\cite{Gasser:1974wd}, and discussed in more detail in Gasser, Hoferichter, Leutwyler and Rusetsky~\cite{Gasser:2015dwa}.
In particular, the most substantial shortcoming of this model is the inability to provide a systematically improvable theoretical uncertainty associated with this model as it is applied in the Regge region where neither low-energy effective field theory or perturbative QCD arguments are applicable.
We have presented an alternative model~\cite{WalkerLoud:2012bg} which is an interpolation between the low and high $Q^2$ regimes, both of which are anchored by rigorous theoretical foundations.  While this model also does not come with a rigorous theoretical uncertainty, because it is interpolating between two controlled regimes, we argue the theoretical uncertainty estimate is more robust and reliable.
The most substantial shortcoming of this model so far is our inability to precisely determine the isovector nucleon magnetic polarizability.
We look forward to more precise experimental measurements of this quantity, which will substantially improve our ability to predict the QED correction to $M_p-M_n$.
This will provide a rigorous test of low-energy QCD predictions that can be confronted with results from lattice QCD.

\bibliographystyle{JHEP}
\bibliography{cottingham}

\end{document}